# Very low bias stress in *n*-type organic single crystal transistors

M. Barra,[1] F. V. Di Girolamo,[1] N. A. Minder,[2] I. Gutiérrez Lezama,[2] Z. Chen,[3] A. Facchetti,[3,4] A. F. Morpurgo,[2,*] A. Cassinese[1,**]

[1]CNR-SPIN and Department of Physics Science, University of Naples, Piazzale Tecchio 80, 80125 Naples, Italy
[2]DPMC and GAP, University of Geneva, 24 Quai Ernest-Ansermet, CH-1211 Geneva, Switzerland
[3]Polyera Corporation, 8045 Lamon Avenue, Skokie, IL 60077 (USA)
[4]Department of Chemistry and Material Research Center, Northwestern University, 2145 Sheridan Road, Evanston, IL 60208 (USA)

**ABSTRACT**

Bias stress effects in *n*-channel organic field-effect transistors (OFETs) are investigated using PDIF-CN$_2$ single-crystal devices with Cytop gate dielectric, both under vacuum and in ambient. We find that the amount of bias stress is very small as compared to all (*p*-channel) OFETs reported in the literature. Stressing the PDIF-CN$_2$ devices by applying 80 V to the gate for up to a week results in a decrease of the source drain current of only ~1% under vacuum and ~10% in air. This remarkable stability of the devices leads to characteristic time constants, extracted by fitting the data with a stretched exponential – that are $\tau \sim 2 \cdot 10^9$ s in air and $\tau \sim 5 \cdot 10^9$ s in vacuum – approximately two orders of magnitude larger than the best values reported previously for *p*-channel OFETs.

Corresponding authors:

[*]alberto.morpurgo@unige.ch

[**]cassinese@na.infn.it

Over the last few years, high-quality *n*-type transistors based on organic semiconductors have been demonstrated[1], with performance close to that of the best *p*-type devices[2], and capable of operating under ambient conditions. Among these new *n*-type materials, Perylene Diimide molecules have received great attention, due to both their self-assembling properties and the tunability of the LUMO level, resulting in highly robust charge transport properties[3]. Currently, *N,N'*-bis(n-alkyl)-(1,7 and 1,6)-dicyanoperylene-3,4:9,10-bis(dicarboximide)s (PDIF-CN$_2$) is probably the most attractive compound. Thin films can be deposited both by evaporation[4] and from solution[5] leading to high carrier mobility. Single crystal devices[6] exhibit outstanding properties, including the largest electron mobility values[7] reported for *n*-type organic FETs and band-like electron transport[8]. With these new molecular materials enabling top quality *n*-type FETs to be realized, it becomes important to explore all aspects of the device electrical characteristics that could limit the device performance.

From a practical point of view, an important issue is the degradation of the field-effect transistor performance under prolonged application of a gate voltage ($V_{GS}$). This phenomenon, known as bias stress[9], consists in the continuous decrease of the drain-source current ($I_{DS}$) when the transistor is driven in the accumulation regime. The bias stress effect has been widely investigated for *p*-channel devices. It was observed that the physical and chemical nature of the dielectric/organic interface plays a major role in determining the magnitude this effect[10,11], but the precise microscopic origin of the phenomenon has remained elusive. Two possible mechanisms have been proposed for *p*-channel (hole accumulation) devices: one scenario attributes the effect to holes that are transferred from the FET channel to localized states in the gate insulator[11]; another scenario invokes hole-assisted generation of protons in the presence of water, with the protons subsequently diffusing into the gate dielectric[10]. Both mechanisms lead to the accumulation of positive charge (protons or holes) in the dielectric, which screens the external applied gate voltage, resulting in a reduction of the source-drain current.

In *n*-channel devices, the phenomenon has so far received little attention[12,13,14], and it is important to start performing systematic experiments both to understand whether comparing *p*- and *n*-type FETs can help elucidating the microscopic origin of this effect, and to find whether bias stress can be minimized in *n*-type organic FETs. In this paper, we explore the bias stress effect in electron transporting PDIF-CN$_2$ single-crystal transistors fabricated with Cytop as the gate dielectric. Cytop was chosen because at its surface no chemical groups are present that act as electron traps[15], and because previous work on *p*-type rubrene single crystal FETs has shown very small hysteresis as well as threshold voltage shift upon sweeping the gate voltage[16,17]. We observe that, upon the application of large gate voltage (up to +80 V), the decrease in source drain current as a function of time, $I_{DS}(t)$, is very small both in vacuum and in air, even when the measurements are protracted for an entire week. Specifically, in vacuum the relative change in current is approximately only 1%, so that after a few days the decrease of the source drain current becomes smaller than changes caused by temperature fluctuations in the environment. In air, the decrease in source-drain current is larger, but even after one week of stressing the total variation of the source-drain current remains less than 10%. Fitting the data using a commonly used stretched exponential dependence[9,10] enables the extraction of the characteristic time scale $\tau$, which we find to exceed $10^9$ s, two orders of magnitude longer than the largest value extracted in the same way in *p*-type devices. We conclude that, remarkably, PDIF-CN$_2$ single crystal FETs on Cytop have significantly better bias-stress characteristics as compared to any other organic transistor investigated in the past.

PDIF-CN$_2$ crystals were grown by physical vapor transport in a stream of Ar gas as reported previously[7]. After the growth, PDIF-CN$_2$ crystals (typically 1 μm thick and several hundreds of microns in length and width) were laminated manually onto a heavily doped Si substrate (acting as gate) covered with 300 nm of thermally grown $SiO_2$ and with a spin-coated Cytop layer (so that the PDIF-$CN_2$ is in contact with Cytop) with prefabricated Ti/Au electrodes. Fig. 1b shows an optical microscope image of a device. The capacitance per unit area of the $SiO_2$/Cytop bilayer was measured with a Agilent Technology 4284A Precision LCR meter, and was found to be between 3 and 5.3 nF/cm² (depending on the Cytop:solvent concentration and spin-coating parameters), in agreement with the values expected from the measured thickness of the Cytop layer (195 - 460 nm). The bias stress effect was investigated both in vacuum (~$10^{-4}$ mbar) and in air, under fixed voltage polarization while measuring $I_{DS}(t)$ on different time scales, ranging from few hours to several days. The devices were operated in the linear regime to ensure that the density of the accumulated charges is uniform throughout the active channel and the phenomenological models commonly employed to describe the threshold voltage shift can be applied to the $I_{DS}(t)$ curves[18]. In a number of cases, we also interrupted stressing to measure the device transfer characteristics and look at the threshold voltage shift induced by the stressing procedure.

Fig. 1 shows typical output and transfer curves recorded in vacuum. The linearity of the $I_{DS}$-$V_{DS}$ characteristics at low $V_{DS}$ (see Fig. 1 a) indicates the absence of any significant contact effect. The threshold voltage ($V_{Th}$) and the field-effect mobility were extracted from the transfer characteristics (Fig. 1b) using the relation

$$I_{DS} = \frac{W}{L}\mu C(V_G - V_{Th})V_{DS}. \qquad (1)$$

The threshold voltage was close to 10V and electron field-effect mobility values in the range $\mu = 2 - 3$ cm²/Vs were obtained in different devices, exhibiting only a small dependence on the gate voltage. The observed behavior and the values extracted for these different quantities are comparable to what has been found in devices fabricated in the same way, in which band-like transport of electrons was recently reported[8]. All devices could be reproducibly and stably operated in air, with the field-effect mobility remaining high under ambient condition, exhibiting a suppression of less than 10% as compared to the values measured in vacuum.

Fig. 2a shows the normalized current ($I_{DS}/I_0$) while stressing the device at $V_{GS} = 80$ V ($V_{DS} = 10$ V) for approximately $10^4$ s in vacuum and in air. Over a period of 2.5 hours of continuous stressing in vacuum, the current decreases by only 0.5 %. In air – while still very small on an absolute level – the current decrease is roughly five times larger: this comparison between the measurements in air and in vacuum under the same stressing condition reveals that, as observed in *p*-channel devices, the presence of ambient gases (such as $O_2$ and $H_2O$) enhances the bias stress effect[10]. The decrease in current can be accounted for by a small shift of the threshold voltage $V_{Th}$ towards more positive values, as seen from the transfer curves in Fig. 2b. The field-effect mobility remains unaffected upon stressing.

Fig. 2c shows a semi-log plot of the current decrease of the same device stressed in vacuum over a period of 3.5 days ($3 \cdot 10^5$ s). After approximately two days, the bias stress effect tends to saturate (note that the time axis is in logarithmic scale). Saturation occurs because, with the bias stress effect being very small, the stress-induced decrease in $I_{DS}$ becomes smaller than fluctuations in the source-drain current caused by external factors affecting the devices. The most significant of these factors is temperature fluctuations of the environment that cause small changes in the carrier mobility (which is temperature dependent[8]). Indeed, we observe features in the $I_{DS}(t)$ curves – the small peaks, marked by the arrows in the inset of Fig. 2c – which occur over a "period" of approximately 24 hours (these effects are not normally observed in bias stress experiments, because the stress-induced change in source-drain current is normally much larger). Irrespective of all these details, the change in current after three days of continuous stressing at 80 V is less than 2 % as compared to the initial value.

In order to quantitatively compare the bias stress in our PDIF-CN$_2$ transistors in vacuum and air, $I_{DS}(t)$ curves recorded during one week were fitted[19] using the stretched exponential model

$$I_{DS}/I_0 = \exp[-(\frac{t}{\tau})^\beta] \qquad (2)$$

as it is commonly done to quantify the bias stress effect in both organic and inorganic transistors[9]. Estimating $\tau$ and $\beta$ also allows us to compare our results with those reported for other bias stress experiments[9,10]. Fig. 3a shows the bias stress in air measured on the same device discussed above. After applying $V_{GS}$ = 80 V and $V_{DS}$ = 10 V for one week, $I_{DS}$ decreased by 9 %. The dashed line is the best fit with the stretched exponential model, from which we extract $\tau$ = 1.9±0.2·10$^9$ s and $\beta$ = 0.29±0.04. In Fig. 3b, the measurement in vacuum is fitted with equation 2 up to 10$^5$ s (where the current decay still dominates, before saturation occurs), in which case the best fit gives $\tau$ = 4.7±0.6·10$^9$ s and $\beta$ = 0.38±0.05. The values of $\beta$ that we extract from the analysis agree with those found in many previous bias-stress experiments on $p$-channel devices, which typically give values between 0.3 and 0.4[9,10]. On the contrary, the values of $\tau$ that we find are very large, which is a result of the very small magnitude of bias stress in our devices. As a term of comparison, the values of $\tau$ found in analogous experiments on $p$-channel FETs are usually in the range of 10$^3$ to 10$^7$ s [9,10,14,20]. The largest value reported so far was for P3HT thin film transistors on HMDS/SiO$_2$ gate dielectrics, where $\tau$ = 4·10$^7$ s,[10] two orders of magnitude lower than in the present case. It is particularly remarkable that such a large value of $\tau$ is observed for an $n$-type OFET, since these devices traditionally have suffered from substantial trapping effects due to the presence of hydroxyl groups, oxygen and water[21].

We now discuss whether the experimental results give some indication as to the microscopic mechanism responsible for the bias stress in our devices. Although the time-dependent decrease in source-drain current was analyzed in terms of a stretched exponential – which is used to describe the effect of ion migration into the gate dielectric[9,10] – an equally good fit to our data with comparable values of $\tau$ and $\beta$ is obtained using a stretched hyperbola

$$I_{DS}/I_0 = [1+(\frac{t}{\tau})^\beta]^{-1}, \qquad (3)$$

predicted by the carrier diffusion scenario[11]. Although this observation is not surprising – because it is known that as long as $t/\tau << 1$, the two functions can describe equally well the time dependence seen in bias stress experiments – it implies that the analysis of the time dependence alone cannot be used to discriminate between the two scenarios. However, an indication as to which of the two mechanisms dominates in our device is given by the differences of the measurements in vacuum and in air. Specifically, having found that in ambient the electron mobility is only ~ 10 % lower than in vacuum indicates that exposure to air does not significantly enhance electron trapping. However, the magnitude of bias stress effect in air is nearly an order of magnitude larger (approximately a factor of 6) than in vacuum. Since trapping increases minimally upon air exposure, the difference in magnitude of bias stress in air and in vacuum suggests that the dominant mechanism is the diffusion of negative ions (presumably OH$^-$ due to the presence of water) into the Cytop layer, which partially screen the applied gate voltage. While this conclusion is physically reasonable in devices based on a polymeric dielectric through which ions can diffuse rather easily, it cannot be readily generalized to other cases, in which carrier diffusion may be the dominant mechanism (in other words, the dominant mechanism for bias stress does not need "universal", i.e. the same for all types of devices).

In conclusion, we have investigated the bias stress effect in *n*-type single-crystal PDIF-CN$_2$ field-effect transistors fabricated on Cytop/SiO$_2$ gate dielectrics. In vacuum, after nearly one week days of stressing at $V_{GS}$ = 80 V, $I_{DS}$ decreases by less than 2 %. The bias stress effect is enhanced in ambient conditions, but remains remarkably small (less than 10% after 1 week stressing in air at $V_{GS}$ = 80 V). This extremely small bias stress effect results in very long time constants $\tau$ of ~ 2·10$^9$ s in air and 4·10$^9$ s in vacuum, much longer (two orders of magnitude) than in the best other organic transistors investigated in the past. This finding is particularly remarkable since *n*-type organic devices are traditionally considered to be more susceptible to trapping and ambient oxidants than *p*-type materials[21].

**ACKNOWLEDGEMENT**:

Financial support from EU FP7 project MAMA Grant Agreement No. 264098, from the Italian Ministry of Research under PRIN 2008 project 2008FSBKKL "*Investigation of n-type organic materials and related devices of interest for electronic applications*", from SNF and from the NCCR MaNEP is gratefully acknowledged.

# FIGURE CAPTIONS

FIG.1: (Color online) a) Output curves ($V_{GS}$ = 0, 10, ... , 80 V) of a PDIF-CN$_2$ single-crystal transistor measured in vacuum. An optical microscope image of one of the device studied is shown in b) – the scale bar is 200 μm long. From the transfer curve measured in the linear regime ($V_{DS}$ = 10 V) we extract a threshold voltage $V_{Th}$ of +10 V (c) and an onset voltage $V_{ON}$ of -4 V (d). The inset of (c) shows the field-effect mobility extracted from equation 1 versus gate voltage $V_{GS}$.

FIG.2: (Color online) a) Normalized source-drain current $I_{DS}(t)/I_0$ of the device whose data are shown in Fig. 1, measured as a function of time in air (blue squares) and in vacuum (green circles) for a 2.5 hour period ($V_{GS}$ = 80 V; $V_{DS}$ = 10 V). (b) Transfer curves of the device measured before (full symbols) and after (empty symbols) bias stress in air (triangles) and in vacuum (circles). The data show that bias stress induced decrease of source drain current is due to a small shift in threshold voltage, with the mobility remaining unaffected. (c) Semi-log plot of the current decrease of the same device stressed in vacuum over a period of 3·10$^5$ s (approximately 3.5 days). After about two days, the current tends to saturate, and small changes of the ambient conditions cause changes in the current that are larger than those induced by stressing. The arrows in the inset of (c) mark the current peaks spaced by approximately 24 hours, which is indicative of the change in the temperature of the environment.

FIG.3: (Color online) Semi-log plots of the normalized current ($I_{DS}(t)/I_0$) measured in air (a) up to 6·10$^5$ s (~ 1 week) while stressing at $V_{GS}$ = 80 V ($V_{DS}$ = 10V) and the same measurements in vacuum (b) up to 10$^5$ s (~28 hours), before the saturation shown in Fig. 2c. The dashed lines are the best fits to the stretched exponential model (Eq.2 in the main text). The time constants $\tau$ extracted from the fits are 1.9±0.2·10$^9$ s in air and 4.7±0.6·10$^9$ s in vacuum. The same data is shown in linear scale in the insets, to show the characteristic time dependence of the bias stress.

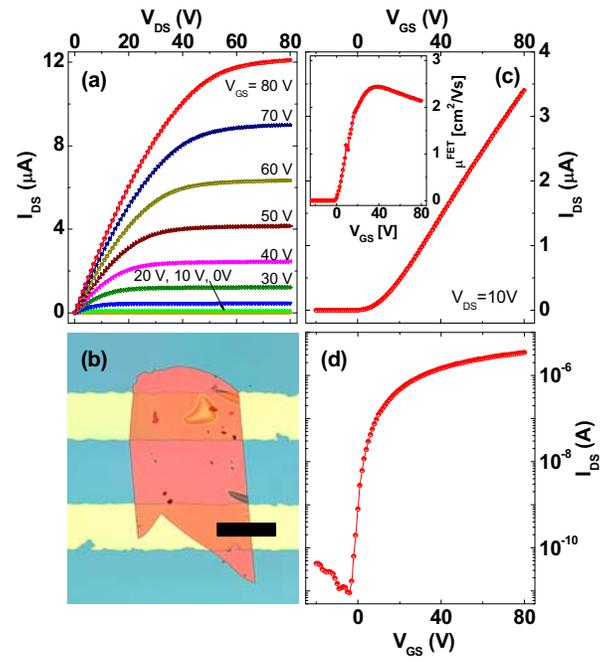

Fig. 1    M. Barra et al.

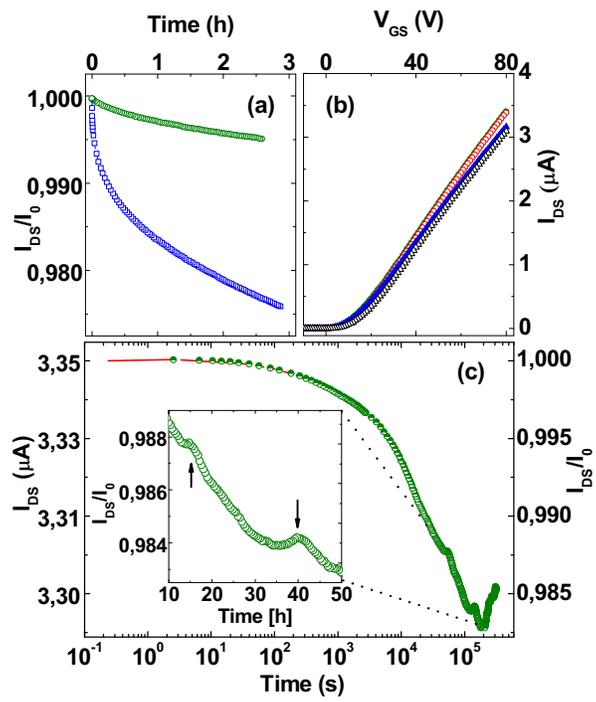

Fig. 2    M. Barra et al.

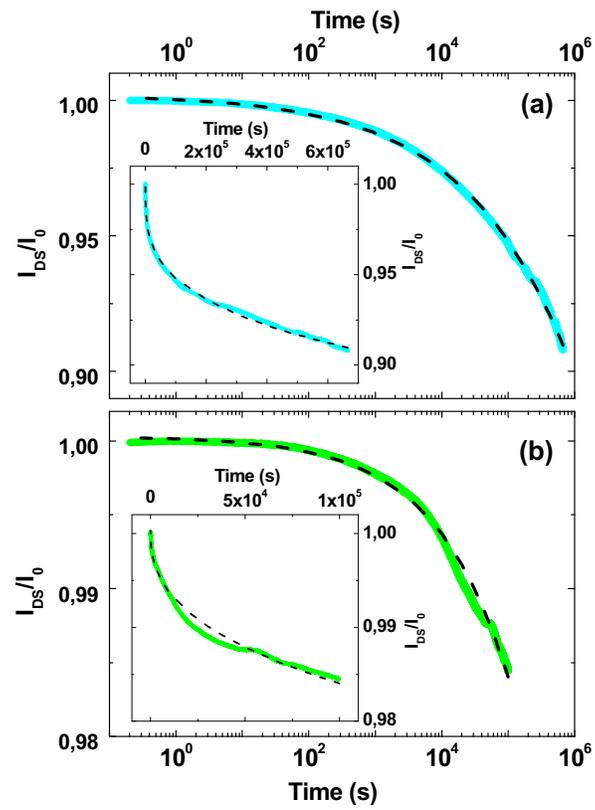

Fig. 3 M. Barra et al.